\documentclass{article}
\usepackage{a4wide}

\title{Inferring genotyping error rates from genotyped trios}

\author{Luke Jostins\thanks{Statistical and Computation Genetics, Wellcome Trust Sanger Institute, Wellcome Trust Genome Campus, Hinxton, Cambs, CB10 1HH}}
\begin{document}
\maketitle

\begin{abstract}
Genotyping errors are known to influence the power of both family-based and case-control studies in the genetics of complex disease. Estimating genotyping error rate in a given dataset can be complex, but when family information is available error rates can be inferred from the patterns of Mendelian inheritance between parents and offspring. I introduce a novel likelihood-based method for calculating error rates from family data, given known allele frequencies. I apply this to an example dataset, demonstrating a low genotyping error rate in genotyping data from a personal genomics company.
\end{abstract}

\section{Introduction}

High-throughput genotyping and sequencing technologies allow affordable genetic studies of a very large number of loci. However, both of these methods have the potential to wrongly infer the genotype of an individual at a particular site. It has been shown that the rate at which these errors occur (the genotyping error rate) can strongly influence the power of a linkage study\cite{pmid11313746}, and moderately influence the power of a case-control study\cite{pmid15475239}.  Estimating error rates is thus an important component of power calculations, as well as in assessing the quality of a dataset.

A popular method of assessing genotyping error rate is to use family relationships between the samples within a dataset. In particular, the rate of impossible inheritance patterns under Mendelian inheritance (the Mendelian error rate) is a commonly used metric. The popular software package PLINK\cite{pmid17701901} has functions to calculate Mendelian error rates in trios per site or per trio. However, the relationship between Mendelian error rates and genotyping error rates is not straightforward, and depends on the allele frequency spectrum of the sites being considered. 

Various methods have been developed to deduce genotyping error rates from Mendelian error rates.  Hao et al\cite{pmid17587543} empirically measured calibration curves between Mendelian errors and genotyping errors, and Saunders et al\cite{pmid12446984} derived an expression for Mendelian error rates in terms of genotyping error rates. However, neither of these methods provide a robust likelihood calculation for observed genotypes for a given error rate. In addition, neither method takes into account the behaviour of the large number of sites that do not contain Mendelian errors, but none the less can still contain information about error rates when coupled with allele frequency measures.

I derive the full likelihood of observed trio genotypes given a genotyping error rate and allele frequencies, and given simplifying assumptions. This can be used to perform maximum likelihood inference of genotyping error rates across sites within a trio or across trios within a site. I apply this method to a trio genotyped by the personal genomics company 23andMe, demonstrating that the trio has a low genotype error rate.

\section{Method}

Here we derive the likelihood of observing a given set genotypes $X$ across $N$ sites in a single trio, though this can equally be applied to the genotypes of $N$ trios at a single site. 

We will denote the site $j \in 1..N$, and the individuals in the trio $i = (o,m,p)$ for offspring, paternal and maternal. We will denote the (unknown) true genotypes $G^{j} = (g_{o}^{j},g_{m}^{j},g_{p}^{j})$, and the observed genotypes  $X^{j} = (x_{o}^{j},x_{m}^{j},x_{p}^{j})$. We will assume a random per-chromosome error rate of $\epsilon$ (the per-diploid genotype error rate is thus approximately $2\epsilon$), and a variant-specific genotype frequency $f^j$. The joint likelihood of true genotype $G^j$ and observed genotype $X^j$ is thus

\begin{equation}
P(X^j, G^j | f^j, \epsilon) =P(x_o^j | g_o^j, \epsilon) P(x_m^j | g_m^j, \epsilon) P(x_p^j | g_p^j, \epsilon) P(g_o^j | g_m^j, g_p^j) P(g_m^j | f^j) P(g_p^j | f^j) 
\end{equation}

Where $P(g_o^j | g_m^j, g_p^j)$ is given by Mendelian transmission and $P(g_i^j | f^j)$ is the frequency of genotype $g_i^j$ under Hardy-Weinberg equilibrium. $P(x_i^j | g_i^j, \epsilon)$ is the error function, with $(1 - \epsilon)^2$ for no error, $(1-\epsilon) \epsilon$ for a heterozygous-to-homozygous or homozygous-to-heterozygous error, and $\epsilon^2$ for a homozygous-to-homozygous error. Note that for heterozygotes there is also a probability  $\epsilon^2$ of a ``double error'' of that doesn't actually change the genotype.

As the true genotype is unknown, we must sum over all possible genotypes:

\begin{equation}
P(X^j | f^j, \epsilon) = \sum_{G^j} P(X^j, G^j | f^j, \epsilon) 
\end{equation}

The overall likelihood for all sites is thus equal to:

\begin{equation}
P(X | \vec{f}, \epsilon) = \prod_{j=1}^{N} \sum_{G^j} P(X^j, G^j | f^j, \epsilon) 
\end{equation}

However, this is computationally expensive to calculate repeatedly. To simplify the calculation, and allow easy recalculation after changing $\epsilon$, we can partition out the joint likelihood into terms that do and do not contain $\epsilon$

\begin{equation}
P(X^j, G^j | f^j, \epsilon) = P(X^j | G^j, \epsilon ) P(G^j | f^j) 
\end{equation}

where 

\begin{eqnarray}
P(X^j | G^j, \epsilon) &=& P(x_o^j | g_o^j, \epsilon) P(x_m^j | g_m^j, \epsilon) P(x_p^j | g_p^j, \epsilon) \\
P(G^j | f^j) &=& P(g_o^j | g_m^j, g_p^j) P(g_m^j | f^j) P(g_p^j | f^j) 
\end{eqnarray}

The overall likelihood can thus be written:

\begin{equation}
P(X | \vec{f}, \epsilon) =  \prod_{j=1}^{N} \left[ P(X^j | G^j = X^j, \epsilon ) P( G^j = X^j | f^j) + \sum_{G^j \neq X^j} P(X^j | G^j, \epsilon ) P(G^j | f^j) \right]
\end{equation}

To simplify this equation, we will assume that only one error occurs (i.e. that all terms in $\epsilon^2$ or greater are negligible). Thus for the first term, where no errors occur $P(X^j = G^j | G^j, \epsilon) \approx (1 - \epsilon)^6$. For genotypes in which one error occurs $P(X^j | G^j, \epsilon)  \approx \epsilon (1-\epsilon)^5$, and for all others $P(X^j | G^j, \epsilon)  \approx 0$. The likelihood then becomes:

\begin{equation}
P(X | \vec{f}, \epsilon) = (1-\epsilon)^{5N}\prod \left[  (1 - \epsilon) P(G^j = X^j | f^j) + \epsilon \sum_{G^j | one\;mutation} P(G^j | f^j) \right] 
\end{equation}

The $P(G^j = X^j | f^j)$ and $\sum P(G^j | f^j)$ terms only need to be calculated once, leaving a relatively easy likelihood calculation to do maximum likelihood estimation on. Point estimates and confidence intervals can then be calculated in the usual way. 

\section{Application}

I applied this method to a trio of individuals genotyped by the personal genomics company 23andMe. This company provides consumers with genotyping and interpretation of their genetic data, while using the data generated to perform case-control association studies for mapping human traits. The company has discovered novel genetic associations\cite{pmid20585627}\cite{pmid21738487}, and replicated many more\cite{pmid21858135}. Low genotyping error rates are critical for both aims, as genotype errors can reduce the power of the case-control analyses, and give consumers false information. 

I used a trio of individuals genotyped by 23andMe, along with allele frequencies from the 60 CEU individuals of the HapMap 3 dataset\cite{pmid20811451}. This allowed assessment of 715 566 variants. The error rate was estimated to be 8.5 x 10$^{-5}$, with a 95\% confidence interal of 6.8-10.2 x 10$^{-5}$, suggesting that genotype errors are rare in this dataset. 

\bibliographystyle{plain}
\bibliography{refs}

\end{document}